# The Impact of Acquisition on Product Quality in the Console Gaming Industry

ECON 191 Topics in Economics, UC Berkeley

Shivam Somani

May 2024

**Abstract**

The console gaming industry, a dominant force in the global entertainment sector, has witnessed a wave of consolidation in recent years, epitomized by Microsoft's high-profile acquisitions of Activision Blizzard and Zenimax. This study investigates the repercussions of such mergers on consumer welfare and innovation within the gaming landscape, focusing on product quality as a key metric. Through a comprehensive analysis employing a difference-in-difference model, the research evaluates the effects of acquisition on game review ratings, drawing from a dataset comprising over 16,000 console games released between 2000 and 2023. The research addresses key assumptions underlying the difference-in-difference methodology, including parallel trends and spillover effects, to ensure the robustness of the findings. The DID results suggest a positive and statistically significant impact of acquisition on game review ratings, when controlling for genre and release year. The study contributes to the literature by offering empirical evidence on the direct consequences of industry consolidation on consumer welfare and competition dynamics within the gaming sector.

# Table of content

# 1. Introduction

## 1.1 Background

The console gaming industry stands as a pillar of the global entertainment landscape, commanding immense popularity and financial significance. With the global gaming revenue surpassing in 2022 reaching $347 billion (Clement, 2024). In recent years, consolidation within the console gaming sector has garnered considerable attention, propelled by high-profile acquisitions and mergers. Notably, Microsoft's acquisitions of major gaming studios such as Activision Blizzard and Zenimax have reshaped the competitive landscape, sparking discussions about their implications for consumer welfare and industry dynamics. Such acquisitions have raised questions about their potential impact on game quality, innovation, and market competitiveness. This was made evident by the UK's markets and competition authority initially blocking the Activision merger (Lomas, 2023) and the FTC authorizing an administrative complaint against the proposed merger (Federal Trade Commission, 2024). Traditionally, vertical mergers like those witnessed in the gaming industry have been viewed through the lens of economic theory, with proponents highlighting potential efficiency gains and synergies while detractors caution against reduced competition and innovation. Understanding the consequences of these mergers is essential not only for industry stakeholders but also for policymakers tasked with ensuring fair competition and consumer protection in the gaming market.

## 1.2 Research Purpose and Organization

Against this backdrop, this study seeks to delve deeper into the impact of acquisitions on game quality within the console gaming industry. By examining empirical evidence derived from a comprehensive dataset spanning two decades of console game releases. The research aims to provide insights into the nuanced effects of industry consolidation on consumer welfare and competition dynamics. Such insights are crucial for informing strategic decision-making and policy formulation in a rapidly evolving gaming landscape. It contributes to the limited existing literature on consolidation in the gaming industry by providing empirical analysis of product quality pre and post merger.

The first section of the paper discusses the existing literature on consolidation in the gaming industry, which contextualizes the study within existing research, highlighting the empirical focus of this research compared to previous theoretical and legal analyses. Moreover the section also provides a summary of prior literature in which a difference-in-difference model was used to ascertain the effects of acquisitions/mergers in other industries. The next section focuses on the research design and strategy outlining the econometric model employed for analysis, explaining its components and how it facilitates the estimation of acquisition effects on review scores. This section also includes the identifying assumptions underlying the difference-in-difference model and how they are addressed to ensure the validity and robustness of the analysis.

The study then delves into data exploration, detailing the dataset used and the variables analyzed to understand the impact of acquisitions on game quality along with providing summary statistics for out control and treatment groups. Finally, the study presents the initial data results and analysis, discussing key findings and observations derived from regression coefficient estimates. It evaluates the significance of the treatment effect and explores potential biases or limitations in the research design.

## 2. Literature Review

### 2.1 Related to consolidation in the gaming industry

My paper will specifically analyze the market for games released on consoles and not the effect of consolidation by Sony, Microsoft & Nintendo on the market for consoles itself as Timothy Derdenger (2010) investigated how technological tying (when a console manufacturer produces software only compatible with its own hardware) affected console price competition and consumer welfare. The study concludes that we shouldn't be too concerned about how these mergers will affect competition in the market for the actual consoles; therefore the focus of this paper was chosen to be empirically determining the effect on consumer welfare in the market for games.

Clayton Alexander and Cahill Maffei approached the issue of consolidation in gaming in their respective journal articles through the lens of antitrust law. They delve into the broader context of vertical mergers in the technology and gaming sectors, with a particular emphasis on antitrust considerations. Alexander observes a concerning trend where first-party studios are increasing their market share, verging on an antitrust violation, though no laws had been breached at the time of his writing in 2020. This observation predates Microsoft's acquisitions of Activision-Blizzard and Zenimax. His analysis employs a theoretical and legal perspective to explore this issue. In contrast, my research will utilize empirical data to examine the impact on product quality, offering a grounded assessment of whether these market shifts detrimentally or positively affect consumer welfare and competition.

Ricard Gil and Frederic Warzynski use data on video game sales and mergers and acquisitions from October 2000 to October 2007. The paper finds that integration leads to higher revenues, more units sold at higher prices, largely due to better release and marketing strategies, rather than game quality. This analysis directly relates to my research by empirically demonstrating the effects of industry consolidation on game performance, hinting at the mechanisms through which consolidation affects the market. While this paper concludes that the difference in performance was due to non-quality related factors, it is also over 10 years old and doesn't represent current market dynamics.

**2.2     Related to using a difference-in-difference model to analyze M&A**

Erdős, Baczur, Kehl, and Farkas (2022) investigate the post-merger price effects in the gasoline retail market using a DiD estimation strategy. Their study provides empirical evidence of significant price increases following the acquisition of two branded gasoline chains. This research emphasizes the necessity of examining product differentiation in evaluating merger effects and therefore influenced the fixed effects chosen for this study, namely the inclusion of genre as a fixed effect. While this study focuses on the impact on prices instead of quality it still highlights the relevance of DiD models in assessing dynamic post-merger outcomes.

Umashankar, Bahadir, and Bharadwaj (2021) investigate the customer satisfaction implications of M&As, providing precedent for using a difference-in-difference model to analyze the effect of mergers on product quality. Their study employs a quasi-experimental DiD analysis to demonstrate the negative impact of M&As on customer satisfaction, outweighing any gains in firm efficiency. By examining the attention-based view of the firm, the authors illustrate how

post-M&A customer dissatisfaction arises from a shift in executive focus away from customers towards financial concerns. This research underscores the importance of considering customer reactions in M&A evaluations and identifies executive attention to customer issues as a crucial factor in mitigating negative outcomes.

Borusyak, Jaravel, and Spiess (2024) present a methodological framework for difference in differences (DiD) designs with staggered treatment adoption and heterogeneous causal effects, offering insights into the analysis of merger and acquisition (M&A) events. Their research addresses the limitations of conventional regression-based estimators in capturing unbiased estimates of relevant parameters, particularly in the presence of treatment-effect heterogeneity. The methodological framework extends to scenarios with time-varying controls, triple-difference designs, and non-binary treatments, offering versatility in studying dynamic post-merger effects. Their research contributes to advancing the methodological toolkit for analyzing the topic of this paper as it provides strategies for how to deal with multiple intervention periods. Incorporating this study into the literature review section provides a comprehensive overview of recent advancements in DiD analysis, highlighting its applicability in studying the multifaceted implications of M&As on various economic and social outcomes.

## 3. Research Design and Strategy

### 3.1 Why a difference-in-difference model

As highlighted in the previous section, the literature to date presents notable gaps, particularly a scarcity of empirical studies focusing on the interplay between game quality and its ownership structure, as well as the broader implications of the gaming industry's consolidation trend on consumer welfare through the lens of game quality. My proposed methodology is designed to investigate whether being acquired has a statistically significant influence on the quality of games, as measured by critic scores through a difference-in-difference model. The decision to prioritize game quality over price as our outcome variable is influenced by the uniform pricing strategy prevalent among AAA video games across different consoles.

In addition to precedent set by the prior literature on analyzing M&As through a difference in difference model, the primary reason for employing a DiD model in place of OLS is to effectively control for all unobserved time-invariant characteristics that could influence the dependent variable such as evolving technology, changes in consumer preferences. This is achieved by examining the difference in changes over time between our treatment group (acquired developers) and our control group (failed acquisitions). It assumes that these unobserved characteristics remain constant over across the two groups, allowing the model to isolate the effect of the intervention.

The DiD method also provides a clear empirical framework for interpreting the causal effect of acquisition by directly measuring the 'treatment effect' as the difference in outcome changes over time between the treatment and control groups given by the interaction term Treatment:PostTreatment.

### 3.2     Choosing a suitable Control

The choice of control group directly influences the ability to isolate the causal effect of acquisition on game quality, therefore selecting an appropriate control group is pivotal for robust research outcomes. A good control group helps mitigate selection bias and other forms of bias inherent in observational studies. By closely matching characteristics of the treatment group, the control group acts as a benchmark for comparison, reducing the risk of biased estimates.

Therefore for this study the control group was chosen to comprise developers which had received acquisition/merger bids but which failed to complete due to negotiations breaking down, not getting board approval or macroeconomic factors causing one of the parties to back out. Therefore our control group would likely be able to account for the anticipation effect of the mergers and would share similar characteristics with the treatment group of acquired developers. The control group also consists of developers similar to acquired studios in terms of market position and size, ensuring comparability between groups. Therefore the majority of the control group is publisher developers similar to Activision and Zenimax. Moreover the control group's timeline aligns closely with the treatment group to account for temporal trends or external shocks that may affect game quality. The list of developers chosen to be the treatment and control groups is attached to the appendix.

### 3.3 Regression Equation and Fixed Effects

Econometric Model: The core of the analysis will be a DiD model to estimate the effect of acquisition on review scores. The regression equation can be specified as:

$$Review\ Rating = \beta_0 + \beta_1 Treatment + \beta_2 Post\ Treatment + \beta_3(Treatment*Post\ Treatment) + X_t Time\text{-}specific + X_i Platform_i + X_j Genre_j + \epsilon$$

Interpretation of Coefficients:
- $\beta_0$ : The intercept representing games developed by non-acquired studios.

- $\beta_1$ : The average change in review scores after acquisition, disregarding the acquisition status.
- $\beta_2$ : The average difference in review scores for games from acquired developers, disregarding the timing relative to acquisition.
- $\beta_3$ : The DiD estimator of interest, representing the additional effect on review scores for games released by developers after being acquired, relative to the expected trend based on other games.
- We also have controls for genre, time specific factors, platform and number of reviews with i, t and j representing a list [0,1,2…] each associated with a specific genre, time period or platform.
    - Genre Fixed Effects: To control for genre-specific biases in ratings.
    - Platform Fixed Effects: To account for platform-specific quality standards and audience expectations.
    - Release Year Fixed Effects: To control for industry-wide trends over time, such as increasing production values or changing consumer preferences.
- $\epsilon$ is the error term.

The coefficient $\beta_3$ directly estimates the causal impact of developer acquisitions on game review scores, answering the research question. A significant positive value would suggest acquisitions improve game quality (as perceived by reviewers), while a significant negative value would suggest the opposite.

Therefore by employing a difference-in-differences analysis, the paper will evaluate the mean quality of games released by acquired studios compared to control mean score of those developed by other major-publisher owned studios with the running variable of time to acquisition. This analysis aims to identify if the acquisition leads to an improvement or decline in game quality relative to games developed by our control group, offering a clear measure of the consolidation's impact over time and the effect of a change in ownership on product quality for acquired studios.

A notable problem in my research design that could result in biases is if acquisitions are not exogenous (e.g., companies are acquired because of anticipated future performance which is how acquisition decisions are made), the estimates could be biased upward. However, given that our control group would also likely experience these effects we can assume that our estimate will provide the actual effect of the acquisition.

### 3.4     Identifying Assumptions for Difference-in-Difference model:

Assumption 1: Parallel Trends

The parallel trends assumption states that in the absence of any acquisition, the trajectory of game ratings for both treated (acquired studios) and control groups (studios which failed to be acquired) would have been parallel over time. The validity of this assumption can be seen by plotting mean review scores against time since acquisition with the confidence interval as seen in figure 1. While this time series plot shows some variance in the difference between the treatment and control groups this is to be expected as individual years might have outliers which performed better or worse than expected. We further test this assumption by creating an event study plot in the data results section later in the paper.

Assumption 2: Spillovers and SUTVA

It's crucial to acknowledge that mergers within the gaming industry might have indirect effects on non-acquired entities due to changes in market dynamics and competitive pressures, potentially violating the SUTVA assumption. However given the practice of uniform pricing and the fact that game development takes place independently across developers it is likely that there are limited spillovers.

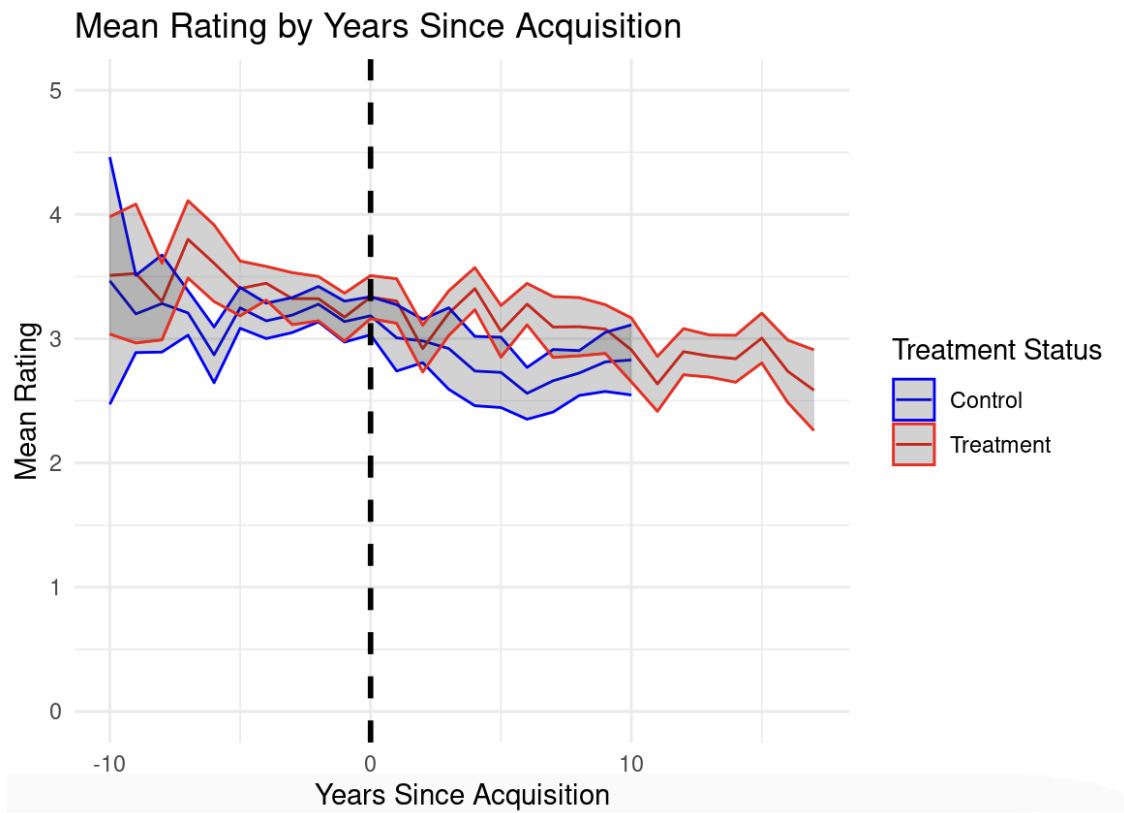

Figure 1: Mean Rating against time with confidence interval

Assumption 3: No anticipation effects

This assumption along with assumption 1 implies that the outcome variable (rating) for untreated observations follow the TWFE model. As acknowledged in Borusyak, Jaravel et al this assumption may require adjustment due to potential anticipation effects. Specifically, it is plausible that management in companies targeted for acquisition might have foreseen these changes, potentially influencing the ratings preemptively. It is likely that our choice of control group mitigates these effects as this selection strategy ensures that any anticipatory behaviors affecting the outcome variable in the treatment group would similarly influence the control group, thereby allowing a more robust and unbiased estimation of the treatment effect. This methodological choice enhances the credibility of the analysis by paralleling the conditions experienced by the treatment group.

## 4.    Data

### 4.1    Data Sources and Data Exploration

The Dataset for this study was acquired from Backloggd a website used to track a user's game collections and allowing them to rate the games they have played. Originally, I was using data from metacritic, a review aggregator, but in 2016 they changed their terms of service not allowing users to scrape the data from their website therefore this alternative data source was found. The data was scraped by Simon Garanin and made available to the public on Kaggle. The dataset includes over 175,000 observations indexing every video game released across all platforms. The data includes multiple variables measuring key attributes for each game including platforms released, release date, genre, developer and number of reviews.

| Variable Name | Explanation |
| --- | --- |
| id | Unique numerical id assigned to each game |
| name | Game title |
| release_year | Release Year |
| rating | Review rating on a scale of 1-5 |
| reviews | Number of reviews |
| platforms | Platforms on which the game was released |
| genres | Genre classification |
| developers | Development Studio(s) responsible for creating the game |
| plays | Number of backloggd users who have played the game |
| playing | Number of backloggd users playing the game |
| wishlist | Number of backloggd users who have wishlisted the game |

For this study we are filtering this dataset to only include games released for the console gaming market composed of consoles released by Sony Playstation, Microsoft Xbox and Nintendo. Moreover, considering the immense technological differences between game development in the 80s or 90s and modern game development only games released post year 2000 will be used in the analysis for this paper. The filtered dataset contains the 16,358 unique games that were released for consoles in the period of 2006-2023.

### 4.2     Creating Variables for Treatment Status

I have created a data frame consisting of console game developers acquired in the period 2006-2023 based on the information from statista along with a dataframe with console developers which failed to be acquired. These data frames have been added to the appendix of the paper. The data frame contains a variable acquired_developers referring to the name if the developer that was acquired along with the acquired_year, the year in which the company was acquired. Using this information I have mutated the original dataset to include the additional variables below.

| Variable Name | Explanation |
| --- | --- |
| Treatment | Dummy Variable denoting whether the developer was or will be acquired |
| PostTreatment | Dummy Variable denoting whether the acquisition has taken place. (1= acquisition has happened) |
| AcquiredYear | Year in which the Developer was acquired. NA for games made by developers which haven't been acquired |
| years_since_acquisition | Release_year - AcquiredYear. NA for games made by developers which haven't been acquired |

### 4.3 Summary Statistics

Table 2 presents the summary statistics for the control and treatment groups in both pre- and post-acquisition periods. The data is organized across four distinct groups: Control Pre-Acquisition, Treatment Pre-Acquisition, Control Post-Acquisition, and Treatment Post-Acquisition. The table provides the mean ratings, the number of observations, and the T-statistics for each group. The table shows that the mean rating of both the control and treatment pre-acquisition is quite similar with the mean for treatment being only 0.08 points higher. This difference does increase in the post treatment period but that is to be expected. The T-statistics for all 4 groups is quite high confirming that the mean differences are statistically significant.

| Group | Mean Rating | Number of Observations | T-Statistic |
|---|---|---|---|
| Control Pre-Acquisition | 2.85 | 364 | 73.49 |
| Treatment Pre-Acquisition | 2.93 | 861 | 112.37 |
| Control Post-Acquisition | 3.11 | 479 | 93.80 |
| Treatment Post-Acquisition | 3.34 | 301 | 82.91 |

Table 2: Summary Statistics

## 5. Results

### 5.1 Regression Results and Analysis

Table 2 gives the regression coefficient estimates for 4 difference-in-difference models with different fixed effects in each. Given that our outcome variable is ratings on a scale of 0-5 and is a proxy for consumer welfare, each coefficient estimate corresponds to that magnitude change in the average rating outcome.

Key Observations

- The key DiD estimator (Treatment:PostTreatment) shows a positive effect in models 1, 2 & 3 however it is only significant in Column 2 controlling for genre and release year, but not platform. This suggests that at least according to the model represented by Column 2 the causal effect of being acquired is a 0.184 increase in expected review rating. This relates directly to my research question as it shows the positive impact being acquired has on product quality due to the additional resources at the disposal of the Developer.

- The positive and significant PostTreatment effects across several specifications may suggest that there are time trends that are increasing ratings over time, separate from the treatment effect.
- A significant constant across all models suggests that the baseline rating is statistically significantly different from zero, likely around 2.85, which is expected in rating data as the data would be centered around the middle. The estimate for column 1 seems inaccurate as ratings are being measured on a scale of 1-5 but the estimate is multiple times larger than the maximum value of 5 at 24.098 and therefore can be dismissed.
- The most reliable estimates likely come from the models with more controls (Column 1 and 2), as they control for more potential confounders. However Column 1 might suffer from over controlling as each of the fixed effects have multiple dummy variables associated with them which might lead to biased estimates. This is backed up by the fact that our constant or intercept estimate is far too large in Column 1.
- The R-squared values range from 0.047 (very low) to 0.405, indicating the proportion of variance in the ratings explained by the model varies significantly depending on the fixed effects included.
- The F-statistics are significant in all models, suggesting that the models are a better fit than an intercept-only model.
- The residual standard error provides an estimate of the standard deviation of the unexplained component of the dependent variable, and it decreases as more controls are added to the model, indicating a better fit.

Table 1: Regression Coefficients with Different Fixed Effects

| | Dependent variable: | | | |
|---|---|---|---|---|
| | | rating | | |
| | Genre, Platform, Release Year | Genre, Release Year | Release Year | No Controls |
| | (1) | (2) | (3) | (4) |
| Treatment | 0.099 | 0.026 | 0.049 | 0.183*** |
| | (0.061) | (0.060) | (0.059) | (0.047) |
| PostTreatment | 0.143* | 0.100 | 0.162** | 0.344*** |
| | (0.075) | (0.075) | (0.067) | (0.054) |
| Treatment:PostTreatment | 0.051 | 0.184** | 0.131 | −0.043 |
| | (0.094) | (0.094) | (0.090) | (0.074) |
| Constant | 24.098 | 2.895*** | 2.917*** | 2.852*** |
| | (21.163) | (0.115) | (0.071) | (0.040) |
| Observations | 1,612 | 1,612 | 1,612 | 1,612 |
| $R^2$ | 0.405 | 0.353 | 0.328 | 0.047 |
| Adjusted $R^2$ | 0.162 | 0.132 | 0.111 | 0.045 |
| Residual Std. Error | 0.640 (df = 1143) | 0.652 (df = 1201) | 0.659 (df = 1218) | 0.683 (df = 1608) |
| F Statistic | 1.664*** (df = 468; 1143) | 1.595*** (df = 410; 1201) | 1.512*** (df = 393; 1218) | 26.356*** (df = 3; 1608) |

Note: *p<0.1; **p<0.05; ***p<0.01

## 5.2 Event Study Plot and proving parallel trends

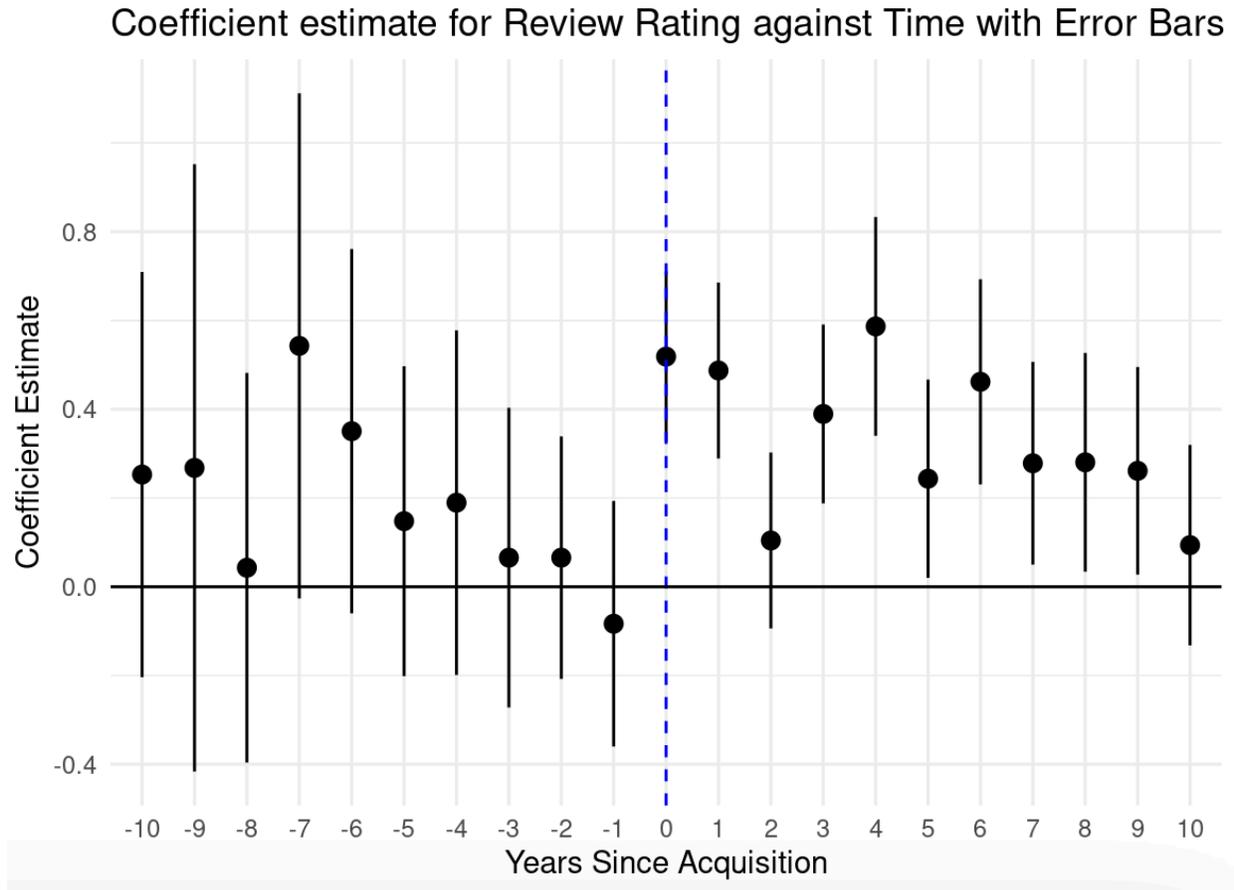

Figure 2: Event Study Plot

The above event study plot provides a comprehensive visual analysis of the changes in game ratings relative to the acquisition timeline. The x-axis represents years relative to the acquisition event, with the blue vertical dashed line at Year 0 indicating the year when the acquisition occurred. The y-axis denotes the coefficient estimates, indicating the variation in game quality (review scores) over time. The error bars represent 95% confidence intervals for each point estimate.

The coefficient estimates preceding the acquisition are relatively stable and consistent with zero, suggesting that the treatment and control groups followed parallel trends prior to the acquisition.

This reinforces the parallel trends assumption, a crucial requirement for difference-in-differences analysis.

Immediately after the acquisition (Years 0 and 1), the coefficient estimates rise slightly and remain positive, with error bars not intersecting zero. This indicates that there is an initial increase in ratings following the acquisition which is statistically significant. However there appears to be a noticeable drop in quality in Year 2, with the coefficient estimate dipping below zero. This decline suggests that game ratings tend to decrease two years post-acquisition, potentially reflecting the transition period where the acquired studio begins implementing strategic changes aligned with the parent company's objectives. It is plausible that this drop coincides with the operational integration of the acquired studio, as it typically takes around two years for the acquisition to be fully realized in practice. This transition often involves management restructuring, cultural shifts, and changes in development priorities, which could initially disrupt production workflows and lead to lower-quality game releases.

After the initial dip, the coefficient estimates gradually recover and stabilize around positive values. However, confidence intervals widen significantly, suggesting increased variability in game quality outcomes over time. The recovery phase may signal that acquired studios adapt to their new environment, capitalizing on additional resources and support.

## 6. Limitations & Potential Biases

While the difference-in-differences (DiD) methodology provides valuable insights into the impact of acquisitions on game quality, there are certain biases and limitations inherent in the research design that warrants consideration. The first source of such bias is in the form of user preferences bias in the primary data source for this study from Backloggd, a platform where users voluntarily rate and review video games. This introduces potential selection bias because the data reflects the preferences and opinions of a specific subset of gaming enthusiasts who are active on the platform. These users may differ in their demographic composition, gaming preferences, or review criteria compared to the broader gaming population. For instance, they might favor certain game genres or studios, which could skew the ratings toward higher scores for games aligning with their preferences. This user-driven bias could impact the internal validity of the study by not fully representing the overall gaming market's sentiments.

Another possible limitation is that the control group, composed of developers that experienced failed acquisition attempts, is relatively small. Finding developers who fit this criterion is challenging because most acquisition bids that fail do not always become public knowledge. As a result, the control group may not fully capture a diverse range of studios that could serve as a comprehensive comparison to those acquired. Additionally, the control group studios that failed to be acquired might differ significantly in strategic direction, financial status, or production capabilities compared to acquired studios. Moreover, we aren't taking into consideration that the merger or acquisition failed due to the acquirer foreseeing a drop in the quality of the games. These differences could influence game quality independently of the acquisition itself, potentially biasing the estimation of the treatment effect.  produced by that developer.

An inherent issue in acquisition research is the potential endogeneity in acquisition decisions. Companies targeted for acquisition often have unique characteristics that could make them more appealing for potential buyers, such as innovative capabilities or strong market positioning. These characteristics might also affect game quality independently of the acquisition, leading to an upward bias in the estimates of the treatment effect.

Lastly, there might be a significant time lag between when an acquisition is announced and when its effects on game quality materialize. Anticipation effects which we have already addressed, where developers adjust their strategies or development practices in expectation of an acquisition, can further complicate causal inference. The DiD model tries to control for anticipation effects, which, if not being fully effective, could distort the estimated impact of the acquisition.

## 7. Conclusion

This study explored the impact of acquisitions on game quality in the console gaming industry using a dataset from Backloggd and a difference-in-differences approach. The analysis found that acquisitions generally result in improved game ratings, although a dip in quality is seen about two years post-acquisition, likely due to the time required for integrating acquired studios and implementing new strategies.

Despite these findings, the study acknowledges limitations, including biases from the dataset's user preferences and a small control group of failed acquisitions. Moreover, acquisition decisions may involve endogeneity and anticipation effects, potentially affecting the treatment effect's accuracy.

Overall, this research contributes valuable insights into the relationship between acquisitions and game quality. While the findings suggest a positive trend, they emphasize the complexities and delays inherent in the acquisition process. Future studies can refine these results by expanding the control group and incorporating additional data sources, providing a more comprehensive understanding of mergers' effects on game quality.

# 9. Appendix

| | Developer | Year |
|---|---|---|
| 1 | Arkane Studios | 2021 |
| 2 | Bethesda Game Studios | 2021 |
| 3 | Bethesda Softworks | 2021 |
| 4 | Activision | 2023 |
| 5 | Infinity Ward | 2023 |
| 6 | Blizzard | 2023 |
| 7 | 4A Games | 2020 |
| 8 | Activision Blizzard | 2023 |
| 9 | ZeniMax Online Studios | 2021 |
| 10 | Bungie | 2022 |
| 11 | Mojang | 2014 |
| 12 | Namco | 2005 |
| 13 | Playdemic | 2021 |
| 14 | Gearbox | 2021 |
| 15 | Codemasters Birmingham | 2020 |
| 16 | Codemasters | 2020 |
| 17 | BioWare | 2007 |
| 18 | BioWare Edmonton | 2007 |
| 19 | BioWare Montreal | 2007 |
| 20 | Jagex | 2020 |
| 21 | Saber Interactive | 2020 |
| 22 | Aspyr Media | 2021 |
| 23 | Riot Games | 2011 |
| 24 | Rare | 2002 |
| 25 | Respawn Entertainment | 2017 |
| 26 | Square Enix Europe | 2022 |
| 27 | Ensemble Studios | 2001 |
| 28 | Insomniac games | 2019 |
| 29 | TT Games | 2007 |
| 30 | Eidos Interactive | 2009 |
| 31 | Ninja Theory | 2018 |
| 32 | Avalanche Studios | 2018 |
| 33 | Pipeworks Studios | 2020 |
| 34 | Ubisoft Reflections | 2006 |
| 35 | Relic Entertainment | 2013 |
| 36 | From Software | 2014 |
| 37 | Digital Extremes | 2014 |
| 38 | Slightly Mad Studios | 2019 |
| 39 | Warhorse Studios | 2019 |
| 40 | CroTeam | 2020 |
| 41 | High Voltage Software | 2020 |
| 42 | Good Shepherd Entertainment | 2021 |
| 43 | Big Ant Studios | 2021 |
| 44 | DotEmu | 2021 |

Table 3: List of Developers in Treatment Group

| | Developer | Year |
|---|---|---|
| 1 | Ubisoft Entertainment | 2016 |
| 2 | Ubisoft Montreal | 2016 |
| 3 | Ubisoft Sao Paulo | 2016 |
| 4 | Ubisoft Paris | 2016 |
| 5 | Ubisoft Quebec | 2016 |
| 6 | Ubisoft Australia | 2016 |
| 7 | Ubisoft Casablanca | 2016 |
| 8 | Ubisoft Singapore | 2016 |
| 9 | Ubisoft San Francisco | 2016 |
| 10 | Ubisoft | 2016 |
| 11 | Ubisoft Kyiv | 2016 |
| 12 | Ubisoft Toronto | 2016 |
| 13 | Ubisoft China | 2016 |
| 14 | Ubisoft Annecy | 2016 |
| 15 | Take Two Interactive | 2008 |
| 16 | 2K Games | 2008 |
| 17 | 2K Sports | 2008 |
| 18 | 2K Play | 2008 |
| 19 | 2K | 2008 |
| 20 | Take-Two Interactive | 2008 |
| 21 | Rockstar North | 2008 |
| 22 | Rockstar Games | 2008 |

Table 4: List of Developers in Control Group